\documentclass[aps,twocolumn,groupedaddress,showpacs]{revtex4}

\usepackage{amsmath}
\usepackage{epsfig}
\begin{document}

\preprint{HEP/123-qed}

\title[Short Title]{Nonlocality of Two-Mode Squeezing with Internal Noise}

\author{Sonja Daffer,$^1$ Krzysztof W$\acute{\mbox{o}}$dkiewicz,$^{1,2}$}

\author{John K. McIver$^1$}%
\affiliation{%
    $^1$Department of Physics and Astronomy,
    University of New Mexico,
    800 Yale Blvd. NE,
    Albuquerque, NM 87131   USA  \\
    $^2$Instytut Fizyki Teoretycznej,
    Uniwersytet Warszawski, Ho$\dot{z}$a 69,
    Warszawa 00-681, Poland
    }%

\date{\today}      
\begin{abstract}
\vspace{.1in} \noindent We examine the quantum states produced
through parametric amplification with internal quantum noise.  The
internal diffusion arises by coupling both modes of light to a
reservoir for the duration of the interaction time. The Wigner
function for the diffused two-mode squeezed state is calculated.
The nonlocality, separability, and purity of these quantum states
of light are discussed. In addition, we conclude by studying the
nonlocality of two other continuous variable states: the Werner
state and the phase-diffused state for two light modes.
\\
\end{abstract}


\maketitle

\section{Introduction}
The two-mode squeezed state is the paradigm of the EPR state for
light modes.  In recent years such states have been experimentally
realized and thus have been used in a number of applications
\cite{mandel1995}. In particular, these states can be used to
demonstrate that quantum mechanics is nonlocal.  Using the
polarization states of the two modes, a violation of Bell's
inequality has been experimentally realized \cite{aspect1982}. In
addition, an experiment based on parity measurements to
demonstrate nonlocality has been proposed \cite{banaszek1998}.
Using parity considerations, a positive everywhere Wigner function
has been shown to be nonlocal.

In particular, by using a phase space representation, nonlocality
for the continuous variable squeezed state can be analyzed.
Because squeezed states can be experimentally demonstrated, a lot
of attention has been given to studying how noise affects these
states. Up to the present, nonlocality for a pure two-mode
squeezed state coupled to an external reservoir has been
investigated \cite{kim2000}. This noise has been introduced into
the state by coupling each mode of the squeezed state to
independent external reservoirs. Alternatively, this can be
interpreted as transmitting the two modes through some noisy
quantum channel.

Finding an exact analytic expression for the Wigner function is
often an operose task.  Gaussian Wigner functions are prevalent
because they are exact solutions to equations which can be easily
dealt with. However, in most cases only numerical solutions exist.
And in some cases analytic solutions exist, but only under certain
approximations. An exact, non-Gaussian solution for the Wigner
function for a nondegenerate parametric oscillator has been
obtained for the steady-state \cite{kheruntsyan2000}. Although the
state is a two-mode squeezed state with internal quantum noise,
the steady state solution exhibits no nonlocal features and,
therefore, is not useful for tests of Bell-type inequalities.

Parametric amplification generates the two-mode squeezed state via
nonlinear interactions in a crystal.  A laser beam passes through
a crystal for some time and the output is two modes of light which
are correlated.  The degree of correlation depends on the
interaction time as well as the strength of the nonlinearity.
Ideally, this process generates a pure quantum state which is
nonseparable.  We will consider the nonideal process where quantum
noise is present inside the crystal.

The purpose of this paper is to study the nonlocal features of the
two-mode squeezed state with internal noise.  We solve the Wigner
function exactly to analyze the dynamics of nonlocality and study
the many features of this quantum state, such as separability and
the behavior in the steady state.
We begin this paper by considering two modes which are coupled via
a diffusive nonlinear crystal.  The Hamiltonian has an interaction
term plus noise terms which couple each mode to a heat bath.  A
linear quantum Fokker-Planck equation is solved for the Gaussian
Wigner function of the state.  An analysis of the steady state
conditions is given. In Section III conditions for purity and
separability of the state are provided.  A discussion of the class
of nonseparable mixed states which exhibit quantum nonlocality
follows. An application to Bell's inequality is given in Section
IV, where the effect of noise on the nonlocality of the state is
discussed. With Section V we conclude by examining two types of
mixed entangled states: the continuous variable Werner state and a
phase-diffused state.
\section{Wigner function.}
The Wigner function for a two-mode squeezed state is well-known to
be
\begin{eqnarray}        \label{eq:w2mss}
    W_{2mss}(\alpha_1,\alpha_2)=\frac{4}{\pi^2} \textrm{exp}[-2 \cosh(2r)(|\alpha_1|^2+|\alpha_2|^2)
    \\
    + 2 \sinh(2r)(\alpha_1 \alpha_2+\alpha_1^\star \alpha_2^\star)]
    \nonumber
\end{eqnarray}
for two coherent modes of light $\alpha_1$ and $\alpha_2$.  The
amount of squeezing of the state is determined by the parameter
$r$ which depends on the nonlinearity of the crystal as well as
the interaction time of the light propagating through the crystal.

In this paper, we derive the Wigner function for a two-mode
squeezed state which has internal noise.  In this case a pure
squeezed state is not produced.  Rather, there is a quantum
diffusion process present during the generation of the squeezed
light.  This is intrinsic quantum noise which is present for the
duration of the interaction time.

The Hamiltonian, in the interaction picture, describing the
process of parametric amplification in the presence of noise is
\begin{eqnarray}
    \hat{H}&=&i \hbar \kappa ( \hat{a}^\dagger_1 \hat{a}^\dagger_2
    -\hat{a}_1 \hat{a}_2  )
    +\sum_{i=1,2} \hbar  (\hat{a}^\dagger_i \hat{\Gamma}_i +
    \hat{a}_i
    \hat{\Gamma}^\dagger_i)
\end{eqnarray}
where $\kappa$ describes the nonlinearity of the crystal. The
parameter $\hat{\Gamma}$ is a reservoir operator which introduces
quantum white noise into the system characterized by the mean
photon number $\bar{n}$.

The equation of motion for the density operator which describes
the quantum state of the two light modes is given by the master
equation
\begin{eqnarray}        \label{eq:master}
    \frac{d \hat{\rho}}{dt}
    &=& \kappa (\hat{a}^\dagger_1 \hat{a}^\dagger_2 \hat{\rho} -
    \hat{\rho}
    \hat{a}_1^\dagger \hat{a}^\dagger_2 - \hat{a}_1 \hat{a}_2
    \hat{\rho}
    +\hat{\rho} \hat{a}_1 \hat{a}_2)       \\
    &+& \sum_{i=1,2}
    \frac{\gamma_i}{2}(\bar{n}_i+1) ( 2 \hat{a}_i
    \hat{\rho} \hat{a}_i^\dagger - \hat{a}^\dagger_i \hat{a}_i  \hat{\rho}
    - \hat{\rho} \hat{a}^\dagger_i  \hat{a}_i)   \nonumber \\
    &+& \sum_{i=1,2}\frac{\gamma_i}{2} \bar{n}_i
    ( 2 \hat{a}^\dagger_i
    \hat{\rho} \hat{a}_i -\hat{a}_i  \hat{a}^\dagger_i \hat{\rho} -
    \hat{\rho} \hat{a}_i \hat{a}^\dagger_i)       \nonumber
\end{eqnarray}
which is obtained by averaging over the reservoir variables.  The
two field modes are coupled to a bath which has a mean photon
number given by $\bar{n}$ and we will assume that the single
photon loss rate for each mode is equal so that
$\gamma_1=\gamma_2$.

The method to convert the above operator equation into a c-number
equation is straightforward with the use of the characteristic
function for the Wigner representation defined as
\begin{equation}        \label{eq:charfunc}
    \chi(\mbox{\boldmath$\beta$})\equiv Tr(\hat{D} \hat{\rho})=e^{-\frac{1}{2}
    {\scriptsize{\mbox{\boldmath$\beta$}} }^\dagger {\textbf{\small V}} {\scriptsize{\mbox{\boldmath$\beta$}} } }
\end{equation}
where
$\mbox{\boldmath$\beta$}=(\beta_1,\beta_1^\star,\beta_2,\beta_2^\star)$
is a four-vector and the operator $\hat{D}$ is the displacement
operator for the two modes
\begin{equation}
    \hat{D}=\hat{D}_1 \hat{D}_2=e^{\beta_1 \hat{a}^\dagger_1-\beta^\star_1 \hat{a}_1}
    e^{\beta_2 \hat{a}^\dagger_2-\beta^\star_2 \hat{a}_2}.
\end{equation}
The matrix $\textbf{V}$ is a 4$\times$4 covariance matrix for the
two modes.

A double Fourier transform of the characteristic function defines
the Wigner function $W(\mbox{\boldmath$\alpha$})$ for the two
modes. Using standard procedures the master equation
(\ref{eq:master}) can be mapped into the following Fokker-Planck
equation for the Wigner function
\begin{equation}
    \frac{\partial}{\partial t} W({\bf x},t)=
    \left[ - A_{ij} \frac{\partial}{\partial x_i} x_j+
    \frac{1}{2} D_{ij} \frac{\partial}{\partial x_i} \frac{\partial}{\partial x_j}
    \right]  W({\bf x},t)
\end{equation}
in terms of the real position and momentum variables.  The
transform is given by $\alpha_1= x_1+i x_2$ and $\alpha_2= x_3+i
x_4$, so that $x_1$ and $x_3$ are position variables and $x_2$ and
$x_4$ are momentum variables. The drift matrix, $A_{ij}$, and the
diffusion matrix, $D_{ij}$, are constant matrices, thus defining a
quantum Ornstein-Uhlenbeck process. The drift matrix is
\begin{equation}
    \textbf{A}= \left(
    \begin{array}{cccc}
    -\frac{\gamma}{2} & 0 & \kappa & 0 \\
    0 & -\frac{\gamma}{2} & 0 & -\kappa \\
    \kappa & 0 & -\frac{\gamma}{2} & 0 \\
    0 & -\kappa & 0 & -\frac{\gamma}{2}
    \end{array}
    \right)
\end{equation}
and the positive-definite diffusion matrix is
\begin{equation}
    \textbf{D}= \frac{\gamma}{4}(2\bar{n}+1)\left(
    \begin{array}{cccc}
    1 & 0 & 0 & 0 \\
    0 & 1 & 0 & 0 \\
    0 & 0 & 1 & 0 \\
    0 & 0 & 0 & 1
    \end{array}
    \right).
\end{equation}

The general solution to a linear, multi-dimensional Fokker-Planck
equation is known and can be solved exactly by the method of
Fourier transform \cite{carmichael1999}. The general Green
function solution for a linear Fokker-Planck equation is a
conditional distribution which is a multi-dimensional Gaussian
having the form
\begin{widetext}
\begin{equation}
    W({\bf x},t|{\bf {x'}},0)=\frac{1}{\sqrt{\textrm{det}\textbf{Q}(t)}}
    \textrm{exp}\left[ -\frac{1}{2}({\bf x}-e^{\textbf{A}t}
    {\bf {x'}})^{\scriptsize\textsf{T}}
    \textbf{Q}^{-1}(t) ({\bf x}-e^{\textbf{A}t} {\bf {x'}})
    \right].
\end{equation}
\end{widetext}
The matrix $\textbf{Q}$ is a time-dependent matrix which depends
on the elements of the diffusion matrix and the eigenvalues of the
drift matrix.

From this solution the unconditional distribution is found through
\begin{equation}
    W({\bf x},t)=\int W({\bf x},t|{\bf {x'}},0) W({\bf {x'}},0) \textrm{d}
    {\bf {x'}}.
\end{equation}
The initial condition is taken to be the two-mode vacuum state
given by
\begin{equation}
    W({\bf x},0)=\left( \frac{2}{\pi} \right)^2 e^{-2(x_1^2+ x_2^2+
    x_3^2+x_4^2)}
\end{equation}
in terms of the real variables.  After integration over the primed
variables and transforming back to the complex variables, we have
the following form for the Wigner function
\begin{equation}        \label{eq:wigner}
    W(\mbox{\boldmath$\alpha$},t)=\left( \frac{2}{\pi} \right)^2
    \frac{1}{h(t)} \hspace{.1in}
    \textrm{exp} \left[- \frac{1}{2} \left( \frac{
    \emph{f}(\alpha_1,\alpha_2,t)
    }{h(t)} \right)  \right]
\end{equation}
where
\begin{eqnarray}
    \emph{f}(\alpha_1,\alpha_2,t)&=&c_1(t) (\alpha_1
    \alpha_1^\star+\alpha_2
    \alpha_2^\star) \\
    &+& c_2(t) (\alpha_1 \alpha_2+\alpha_1^\star \alpha_2^\star)
    \nonumber
\end{eqnarray}
with time-dependent functions
\begin{eqnarray}        \label{eq:cparameter}
    c_1&=&2(e^{-p_2}+e^{-p_1})  \nonumber \\
    &+&(2\bar{n}+1)(p_1+p_2)
    \left( \frac{1-e^{-p_1}}{p_1} + \frac{1-e^{-p_2}}{p_2} \right) \nonumber \\
    c_2&=&-2(e^{-p_2}-e^{-p_1})  \\
    &+&(2\bar{n}+1)(p_1+p_2)\left( \frac{1-e^{-p_1}}{p_1} - \frac{1-e^{-p_2}}{p_2}
    \right),
    \nonumber
\end{eqnarray}
and
\begin{eqnarray}     \label{eq:hparameter}
    h = \left[ e^{-p_1}
           + (2\bar{n}+1)\left( \frac{p_1+p_2}{2} \right) \left( \frac{1-e^{-p_1}}{p_1} \right)\right] \\
    \times \left[ e^{-p_2}
           + (2\bar{n}+1)\left( \frac{p_1+p_2}{2} \right) \left( \frac{1-e^{-p_2}}{p_2} \right)
           \right] \nonumber.
\end{eqnarray}
The dimensionless time parameters $p_1=d+2r$ and $p_2=d-2r$ have
been used. We call $d=\gamma t$ the diffusion parameter and
$r=\kappa t $ the squeezing parameter.  It should be emphasized
that $c_1=c_1(t), c_2=c_2(t),$ and $h=h(t)$ and, throughout this
paper, we occasionally omit the function's dependent variables for
simplicity.

The eigenvalues of the drift matrix are doubly degenerate and are
\begin{equation}
    \lambda_{1,2}=-\frac{1}{2}(\gamma+2 \kappa),
    \hspace{.2in}
    \lambda_{3,4}=-\frac{1}{2}(\gamma-2 \kappa).
\end{equation}
The condition for a steady state solution to exist is that the
eigenvalues of the drift matrix all have negative real parts.
Thus, the parameters $p_1$ and $p_2$ alone determine whether a
steady state solution for the Wigner function exists.  If
$\frac{\gamma}{2 \kappa}>1$ then a steady-state solution exists.
The steady state Wigner function is a squeezed thermal state given
by
\begin{equation}
    W=
    \frac{\textrm{exp} \left[- \frac{2}{2\bar{n}+1} \left(
    |\alpha_1|^2+|\alpha_2|^2 -\frac{2\kappa}{\gamma} \left(
    \alpha_1\alpha_2+\alpha_1^\star \alpha_2^\star \right)
    \right)  \right] } { \left( \pi/2 \right)^2
    (2\bar{n}+1)^2/\left(1-(\frac{2\kappa}{\gamma})^2\right)}.
\end{equation}
This steady state solution is undefined if $2\kappa=\gamma$.  If
$\kappa=0$ the steady state solution is a thermal state.  For the
case of a pure two-mode squeezed state, $\gamma=0$, there is no
steady state solution because as the interaction time approaches
infinity the Wigner function approaches an EPR state with perfect
correlations.
\section{Purity and Separability of Gaussian States}
\subsection{Purity}
In terms of the Wigner representation, a general two-mode Gaussian
state may be written as
\begin{equation}
    W(\mbox{\boldmath$\alpha$})=\frac{\sqrt{\textrm{det}\textbf{W}}}{\pi^2}
    e^{-\frac{1}{2} {\scriptsize{\mbox{\boldmath$\alpha$}} }^\dagger {\textbf{\small W}} {\scriptsize{\mbox{\boldmath$\alpha$}} }}.
\end{equation}
For the state in Eq. (\ref{eq:wigner}), $\textbf{W}$ takes the
simple form
\begin{equation}
\textbf{W}=\frac{1}{2h}\left(
\begin{array}{cccc}
c_1 & 0 & 0 & c_2 \\
0 & c_1 & c_2 & 0 \\
0 & c_2 & c_1 & 0 \\
c_2 & 0 & 0 & c_1
\end{array}
\right).
\end{equation}
The $\textbf{V}$ matrix can be obtained from the $\textbf{W}$
matrix through the relation between the Weyl-Wigner characteristic
function and the Wigner function.  They are related by a Fourier
transform which leads to
\begin{equation}
    \textbf{W}=\textbf{E} \textbf{V}^{-1} \textbf{E}
\end{equation}
where the matrix $\textbf{E}$=diag[1,-1,1,-1]. From this we have
that
\begin{equation}
\textbf{V}= \frac{1}{\sqrt{\textrm{det}\textbf{W}}} \textbf{W}
\end{equation}
so that both matrices differ only by a factor.

The form of the covariance matrix is
\begin{equation}     \label{eq:vmatrix}
\textbf{V}=\left(
\begin{array}{cccc}
N+\frac{1}{2} & 0 & 0 & M \\
0 & N+\frac{1}{2} & M & 0 \\
0 & M & N+\frac{1}{2} & 0 \\
M & 0 & 0 & N+\frac{1}{2}
\end{array}
\right).
\end{equation}
with the identification $N+1/2=2hc_1/(c_1^2-c_2^2)$ and
$M=2hc_2/(c_1^2-c_2^2)$.  $N$ quantifies the correlation $\langle
\hat{a}^\dagger_i \hat{a}_i \rangle$ for mode $i$, while $M$
quantitifies the correlation between modes related to squeezing
$\langle \hat{a}_1 \hat{a}_2 \rangle$.

Gaussian operators which are projectors represent pure states
\cite{englert2002}.  The condition for a Gaussian Wigner function
to represent a pure state may be written concisely as
$\sqrt{\textrm{det} \textbf{W}}=4$. The Gaussian operator
corresponding to Eq.(\ref{eq:wigner}) is a projector provided
\begin{equation}
    c_1(t)^2-c_2(t)^2=16h(t)^2.
\end{equation}
From this condition, it is established that only one pure state
exists, namely, that for which both $r\neq0$ and $\gamma=0$. This
corresponds to a squeezed state which is generated in the absence
of noise. The presence of the internal quantum noise always
produces a mixed state. The resulting mixed states may be
separable or nonseparable, depending on the parameters of the
model.

\subsection{Separability}
A separability criterion for two-mode Gaussian states has been
established which relies on the partial transposition map acting
on the two-party state \cite{simon2000}.  It has been shown that
this criterion is equivalent to determining that the quantum state
is P-representable.  A two-mode Gaussian state is P-representable,
and hence, separable, if and only if
\begin{equation}        \label{eq:sepcondition}
    \textbf{V}-\frac{1}{2} \textbf{I} \geq 0
\end{equation}
where \textbf{V} is the covariance matrix for the characteristic
function found in Eq.(\ref{eq:charfunc}) which takes the special
form of Eq.(\ref{eq:vmatrix}).  The separability condition is also
equivalent to the condition that the matrix elements of
(\ref{eq:vmatrix}) satisfy $N \geq M$.

The state in Eq. (\ref{eq:wigner}) is separable if and only if the
eigenvalues of Eq.(\ref{eq:sepcondition}) are nonnegative. There
are four eigenvalues which are doubly degenerate. They are
\begin{eqnarray}
    e_{1,2}&=&\frac{(1-e^{-p_2})(d\bar{n}+r)}{p_2} \nonumber \\
    e_{3,4}&=&\frac{(1-e^{-p_1})(d\bar{n}-r)}{p_1} .
\end{eqnarray}
The first eigenvalue has the property that the sign$(e_{1,2})$ is
positive for all parameter space.  It is not important for
establishing the nonseparability of the state.  The second
eigenvalue determines the nonseparability of the state. For
certain regions of the parameter space it is negative, while in
other regions it is positive.  The negativity of $e_{3,4}$ may be
summarized as sign$(e_{3,4})$=sign$(d\bar{n}-r)$.  Hence, the
state is nonseparable if and only if
\begin{equation}
    r > d \bar{n}.
\end{equation}
This is the requirement that the squeezing parameter be greater
than the product of the single-photon loss rate described by the
internal diffusion parameter and the mean photon number of the
reservoir. Fig. 1 shows a parametric plot for varying mean photon
number $\bar{n}$. States which lie on or above the solid line are
separable states. The dots show states which have $\bar{n}=0$ for
different values of the diffusion parameter $d$. We immediately
see that if $\bar{n}=0$ then $e_2<0$ for all parameter space so
that the system is never separable. Therefore, one may conclude
that this system is entangled for all of parameter space when
coupled to a zero-temperature heat bath. This is interesting
because the diffusion parameter may be made as large as desired
and still the state is nonseparable, even as the diffusion
approaches infinity.
\begin{figure}[htbp]
    \centerline{\scalebox{0.7}{\includegraphics{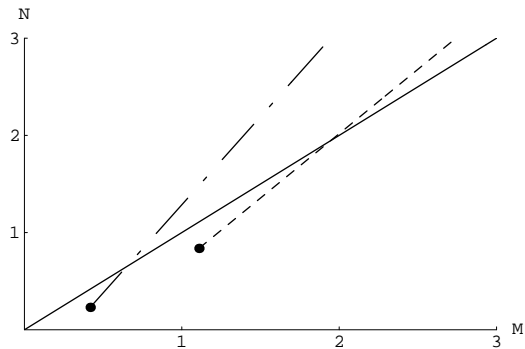}}}
    \caption[short caption.]{A parameteric plot shows the states which are separable.
    States which lie on or above the solid line are separable
    states. The dash-dotted line is for $d=5$ while the dashed line
    is for $d=2.5$.  The two dots correspond to $\bar{n}$=0.  The mean photon number
    increases from 0 to 10 along both lines.  The squeezing parameter is fixed at $r=1.5$. }
\end{figure}
The nonseparable states may be further classified into two sets:
those which exhibit quantum nonlocality and those which do not.
\section{Nonseparable mixed states exhibiting quantum nonlocality}
If the two-mode squeezed state is generated with internal noise
present then the resulting state will be a mixed state.  Although
all pure entangled states violate some Bell inequality, it is not
clear in general which mixed entangled states will do so.  If no
noise is present then there always exists, for a fixed $r$, some
range of values for $J$ which will correspond to a nonlocal state.
The presence of the diffusion parameter will destroy the nonlocal
features until some critical value for d is reached such that the
state goes from nonlocal to classically correlated.  Thus, in
addition to the pure states ($d=0$) which are nonlocal, there is a
set of mixed states ($d\neq0$) which is also nonlocal.

To study the nonlocal properties of the state (\ref{eq:wigner}) we
use the formalism of parity operator correlations developed in
\cite{banaszek1998}.  The Wigner function is related to the
expectation value of a displaced parity operator in phase space
given by $\Pi(\alpha_1,\alpha_2)=\langle
\hat{D}(\alpha_1)(-1)^{\hat{n}_1} \hat{D}^\dagger(\alpha_1)
\otimes \hat{D}(\alpha_2)(-1)^{\hat{n}_2}
\hat{D}^\dagger(\alpha_2) \rangle $. This expectation value
defines a correlation function
\begin{equation}
    \Pi(\alpha_1,\alpha_2)=\left( \frac{\pi}{2} \right)^2
    W(\alpha_1,\alpha_2)
\end{equation}
for parity measurements of mode 1 and mode 2, which is
proportional to the two-mode Wigner function.  In analogy with the
case of dichotomic spin states, a Bell combination may be written,
using correlations between parity measurements, as
\begin{equation}        \label{eq:bellcombo}
    \emph{B}=\Pi(0,0)+\Pi(\sqrt{\emph{J}},0)+
    \Pi(0,-\sqrt{\emph{J}})-\Pi(\sqrt{\emph{J}},-\sqrt{\emph{J}}).
\end{equation}

For a two-mode squeezed state with internal noise the combination
$\emph{B}$ will be a function of the interaction time as well as
the squeezing and noise parameters.  The Bell combination is
\begin{align}
    \emph{B}(\emph{J},t) = \frac{1}{\emph{h}(t)}
    \Biggl[
    1 + 2 \hspace{.05in} \textrm{exp} \Biggr. &\left( -\emph{J}\hspace{.05in} \frac{c_1(t)}{2 h(t)}
    \right) \\
     - \hspace{.05in} \textrm{exp} &\Biggl.\left(
    -\emph{J} \hspace{.05in}\frac{c_1(t)+c_2(t)}{h(t)} \right)
    \Biggr]   \nonumber
\end{align}
where the functions $c_1,c_2,$ and $h$ are given by
Eqs.(\ref{eq:cparameter},\ref{eq:hparameter}) which are implicitly
functions of time, and $\emph{J}$ is the square of the coherent
amplitude of the light mode.

A search algorithm finds a local maximum for $\emph{B}$ and
returns the values of the free parameters.  The function
$\emph{B}$ has a maximum of 2.19 with values $\emph{J}=.01, r=1.5,
d=0$, and $\bar{n}=0$.  As expected, the Bell combination is most
nonlocal when there is no noise present and the squeezed state is
a pure two-mode squeezed state.  In this case, one finds a
stronger violation for large r and small intensity $J$.  To see
how the internal noise affects the nonlocality of the state we set
$r=1.5$ and let $J$ and $d$ vary in Fig.2.
\begin{figure}[htbp]
    \includegraphics{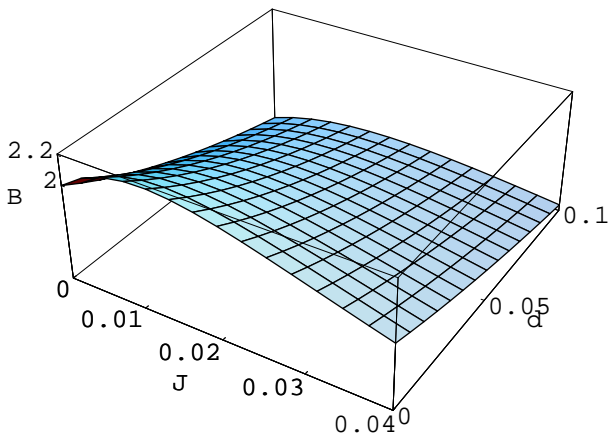}
    \caption[short caption.]{The Bell combination is plotted as a
    function of the coherent amplitude, $J$, for the two field modes and
    the diffusion parameter, $d$, with the squeezing parameter fixed at
    $r=1.5$
    and $\bar{n}=0$.}
\end{figure}
It is clear that the diffusion parameter destroys the nonlocal
features of the state.  Fig. 3 shows a cross-section of the three
dimensional graph of Fig. 2 for $J=.01$.  The Bell combination is
greater than 2 for small values of the diffusion parameter, $d$.
On a large scale, we find that as d increases the Bell combination
decreases, reaches a minimum, and then increases to approach 2
from below.
\begin{figure}[htbp]
    \centerline{\scalebox{0.9}{\includegraphics{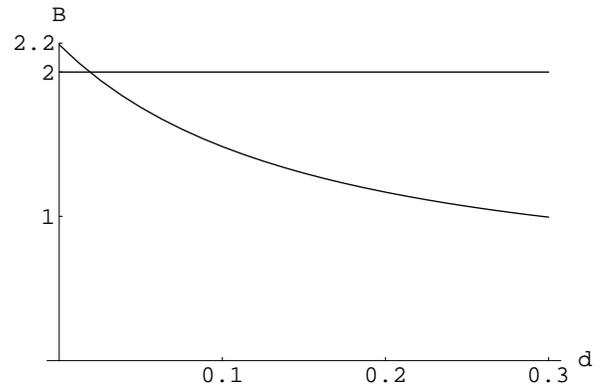}}}
    \caption[short caption.]{The Bell combination is plotted as a
    function of the diffusion parameter, $d$, with the squeezing parameter
    fixed at $r=1.5,\bar{n}=0$, and $J=.01$}
\end{figure}

\section{Continuous Variable Werner States}

Using the two-mode squeezed state of Eq.(\ref{eq:w2mss}), a class
of nonseparable, non-Gaussian, mixed states can be explored.  We
will consider two cases: a convex combination of a pure entangled
state with an uncorrelated mixed state, and a convex combination
of a pure entangled state with a classically correlated state. The
first case is the analogue of the Werner state for continuous
variables.

Werner \cite{werner1989} investigated the state
\begin{equation}
    \hat{\rho}=  p | \Psi \rangle \langle \Psi | + \frac{1-p}{\textrm{d}^2} \hat{I}_A \otimes
    \hat{I}_B
\end{equation}
for $0<p<1$ which is a convex combination of a maximally entangled
state $| \Psi \rangle =\frac{1}{\sqrt{\textrm{d}}}
\sum_{i=1}^{\textrm{d}} |i\rangle_A \otimes |i\rangle_B$ for
finite dimension $\textrm{d}^2$ with a maximally mixed state
obtained by a partial trace over the maximally entangled state.
It was shown that such a state is nonseparable if and only if
\begin{equation}   \label{eq:wernercondition}
    p>\frac{1}{1+\textrm{d}}.
\end{equation}

\subsection{Nonlocality of the continuous variable Werner state}
A natural extension of a Werner state for infinite dimensions is a
convex combination of a maximally entangled state with the
maximally mixed state
\begin{equation}   \label{eq:cvwernerstate}
    \hat{\rho}_W=  p \hat{\rho}^{AB}_{2mss} + (1-p)  \hat{\rho}^{A}_{T} \otimes
     \hat{\rho}^{B}_{T}
\end{equation}
for two modes of light, A and B. This state is analogous to the
Werner state when the squeezing parameter r is infinite.   The
continuous variable phase-space representation of this state is
given by
\begin{equation}      \label{eq:cvwigner}
    W_{W}=  p W_{2mss}(\alpha_1,\alpha_2) +(1-p) W_{T}(\alpha_1)W_{T}(\alpha_2)
\end{equation}
where $W_{T}(\alpha)$ is the marginal Wigner function obtained by
the integration
\begin{equation}
    W_{T}(\alpha)= \int W_{2mss}(\alpha_1,\alpha_2) d^2 \alpha_2=
    \frac{2}{\pi} \frac{e^{-\frac{-2(|\alpha|^2)}{\cosh(2r)}}}{\cosh(2r)}
\end{equation}
over the variables of one mode.  When r is infinite, the Wigner
function $W_{2mss}$ is the EPR state and the function
$W_{T}(\alpha)$ has infinite variance representing a maximally
mixed state of knowledge.  Note that Eq.(\ref{eq:cvwigner}) is not
a Gaussian operator, but rather, it is a linear superposition of
Gaussian operators.  The separability conditions for a linear
superposition of two Gaussian operators have not yet been
established. However, it has been shown that the state
$\hat{\rho}_W$ is nonseparable if $p>0$ \cite{mista2002}. This is
the limit of the finite dimensional case,
Eq.(\ref{eq:wernercondition}), as the dimension of the Hilbert
space, d, approaches infinity.

A test of nonlocality for the state $W_{W}$ in phase space can be
performed using the same Bell combination given by
Eq.(\ref{eq:bellcombo}).  It is found that the state remains
nonlocal for a region of mixtures lying between $p=.9$ and $1$.
Adding a mixed state to the entangled state rapidly degrades the
nonlocal features.
\begin{figure}[htbp]
\centerline{\scalebox{1.0}{\includegraphics{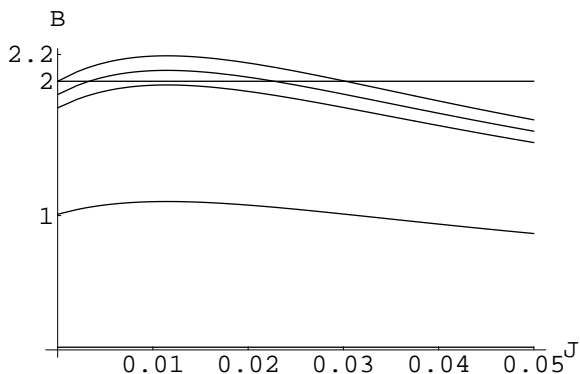}}}
\caption[short caption.] {\small {The Bell combination is plotted
for the continuous variable Werner state.  The squeezing parameter
is fixed at $r=1.5$. The curves are shown for the following values
of mixture, p, from highest maximum value to lowest maximum value:
$p=1$ (top-most curve), $p=.95, p=.9, p=.5$, and $p=0$
(bottom-most curve). } }
\end{figure}
\subsection{Nonlocality of the phase-diffused state}
In addition to the Werner state of Eq.(\ref{eq:cvwernerstate}) for
continuous variables, a different convex combination has some
interesting properties.  Let us consider the state
\begin{equation}     \label{eq:wernertypestate}
    W = p W_{2mss} + (1-p) \overline{W}
\end{equation}
which is a convex combination of a two-mode squeezed state with a
phase-averaged state.  Rather than integrating over one mode of
light, an integration over the phases of both modes is performed.
The state
\begin{equation}        \label{eq:phaseaveragedstate}
    \overline{W}= \int_0^{2 \pi}  \int_0^{2 \pi} W_{2mss} \frac{d \phi_1}{2 \pi} \frac{d \phi_2}{2
    \pi}
\end{equation}
describes phase diffusion of two light modes such that each mode
has a completely random phase.  The integral of
Eq.(\ref{eq:phaseaveragedstate}) leads to a phase-averaged Wigner
function given by
\begin{equation}
    \overline{W}= \frac{4}{\pi^2}e^{-2 \cosh{(2r)}(|\alpha_1|^2+|\alpha_2|^2)}
    I_0(4|\alpha_1||\alpha_2|\sinh{(2r)})
\end{equation}
where $I_0$ is the modified Bessel function of the first kind. In
contrast to the Werner state, this state does not factorize into a
product of functions containing each mode. The state
$\overline{W}$ is a classically correlated state. The state of
Eq.(\ref{eq:wernertypestate}) now being considered is a mixture of
a state containing quantum correlations and a state containing
classical correlations.  The nonlocality of this state is examined
by considering an expansion for small intensity, $J$, of the Bell
combination given by Eq.(\ref{eq:bellcombo}).  This expansion
\begin{equation}
    B=2+4Jp \sinh(2s)+O(J^2)
\end{equation}
reveals that the state (\ref{eq:wernertypestate}) is nonlocal for
all $p>0$.  This is seen in Fig.(5) which shows that only the case
$p=0$ washes out the nonlocal features completely. Thus, we have
presented a classically correlated state which, when mixed with
even the smallest amount of a quantum correlated state, shows
nonlocal features.
    \begin{figure}[htbp]
    \centerline{\scalebox{1.0}{\includegraphics{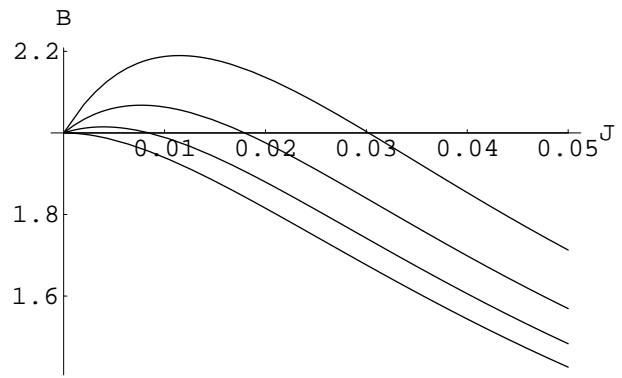}}}
    \caption[short caption.]
    {\small {The Bell combination is plotted for the Werner-type
    state $p W_{2mss}+(1-p)\overline{W}$.  The squeezing parameter is
    fixed at $r=1.5$.  The top-most curve shows the pure two-mode
    squeezed state obtained for $p=1$.  The remaining curves, from the greatest maximum value
    to the lowest maximum value, show
    $p=.5, p=.2$, and $p=0$ (no violation).} }
    \end{figure}

\section{Conclusion}
In this paper we examined the two-mode squeezed state with
internal quantum noise.  The Fokker-Planck equation was solved to
obtain an exact solution for the Wigner function.  The
steady-state and nonsteady-state regimes were explored.  We found
that nonlocal features of this state exist in the regime where no
steady-state solution exists.

The parameter space contains separable and nonseparable states,
both mixed and pure. Only one pure state exists which is
nonseparable. Nearly all states in the parameter space are mixed
states. Using the Gaussian Wigner function for the state, the
separability criterion was determined to identify those mixed
states which are nonseparable. There exists a set of nonseparable
mixed states for which the quantum diffusion can be large
$0<d<\infty$.

Using the Wigner function, a test of nonlocality in phase space
determined the effect of internal quantum noise on Bell's
inequality. As expected, the noise parameters reduce the
nonlocality of the state.  But there is still a region of mixed
entangled states which are nonlocal.  To further study
nonseparable mixed states exhibiting nonlocality, we investigated
Werner-type states for continuous variables.  We found that mixing
the two-mode squeezed state with a product of two thermal states
destroys the nonlocal features state.  Additionally, we found that
mixing the pure two-mode squeezed state with a phase-diffused,
classically correlated state did not destroy the nonlocal
features, regardless of the amount of mixing.  With this, we have
provided an example of a classically correlated state which, when
mixed with even the smallest amount of pure entangled state, is
nonlocal.

\begin{acknowledgments}
This work was partially supported by a KBN grant No. 2PO3B 02123
and the European Commission through the Research Training Network
QUEST.\\
\end{acknowledgments}


\end{document}